\documentclass[11pt]{article}

\usepackage{amsmath}
\usepackage{graphicx}
\usepackage{amsfonts}
\usepackage{amssymb}
\usepackage{epsfig}
\usepackage{color}
\usepackage{psfrag}
\usepackage{epstopdf}

\newcommand{\EQ}{\begin{equation}}
\newcommand{\EN}{\end{equation}}
\newcommand{\be}{\begin{equation}}
\newcommand{\ee}{\end{equation}}
\newcommand{\bea}{\begin{eqnarray}}
\newcommand{\eea}{\end{eqnarray}}

\begin{document} \setcounter{page}{0}
\topmargin 0pt
\oddsidemargin 5mm
\renewcommand{\thefootnote}{\arabic{footnote}}
\newpage
\setcounter{page}{0}
\topmargin 0pt
\oddsidemargin 5mm
\renewcommand{\thefootnote}{\arabic{footnote}}
\newpage
\begin{titlepage}
%\begin{flushright}
%SISSA 02/2014/FISI \\
%DFTT 9/2007
%\end{flushright}
\vspace{0.5cm}
\begin{center}
{\large {\bf Interface in presence of a wall. Results from field theory}}\\
%{\bf Results from field theory}}\\
\vspace{1.8cm}
{\large Gesualdo Delfino$^{1,2}$, Marianna Sorba$^{1,2}$ and Alessio Squarcini$^{3,4}$}\\
\vspace{0.5cm}
{\em $^1$SISSA -- Via Bonomea 265, 34136 Trieste, Italy}\\
{\em $^2$INFN sezione di Trieste, 34100 Trieste, Italy}\\
%{\em $^3$Institute for Theoretical Physics, RWTH Aachen University, 52056 Aachen, Germany}\\
{\em $^3$Max-Planck-Institut f\"ur Intelligente Systeme, 
Heisenbergstr. 3, D-70569, Stuttgart, Germany}\\
{\em $^4$IV. Institut f\"ur Theoretische 
Physik, Universit\"at Stuttgart, Pfaffenwaldring 57, D-70569 
Stuttgart, Germany}\\
\end{center}
\vspace{1.2cm}

\renewcommand{\thefootnote}{\arabic{footnote}}
\setcounter{footnote}{0}

\begin{abstract}
\noindent
We consider three-dimensional statistical systems at phase coexistence in the half-volume with boundary conditions leading to the presence of an interface. Working slightly below the critical temperature, where universal properties emerge, we show how the problem can be studied analytically from first principles, starting from the degrees of freedom (particle modes) of the bulk field theory. After deriving the passage probability of the interface and the order parameter profile in the regime in which the interface is not bound to the wall, we show how the theory accounts at the fundamental level also for the binding transition and its key parameter.
\end{abstract}
\end{titlepage}

\newpage
\section{Introduction}
An important problem in the theory of statistical systems close to criticality is that of providing a fundamental treatment of phenomena involving different length scales. The divergence of the correlation length $\xi$ as the critical temperature $T_c$ is approched is at the origin of universality, namely the existence of quantities such as critical exponents whose values only depend on global properties (internal symmetries and space dimensionality). Field theory then emerges as the natural framework for the quantitative study of universality classes (see e.g. \cite{ID,Zinn}). In particular, the scaling dimensions of the fields, which determine the critical exponents, are related to the behavior of correlation functions at distances much smaller than $\xi$. On the other hand, below $T_c$, in a system with discrete internal symmetry, suitable boundary conditions lead to the presence of an interface separating two coexisting phases. The phenomenon requires a length scale $R$ -- the li
 near size of the interface -- which is much larger than $\xi$, since on shorter scales bulk fluctuations do not allow the emergence of the two distinct phases. There is no doubt that slightly below $T_c$ the full description of the system with the interface should be obtained supplementing with the required boundary conditions the field theory of the bulk (i.e. homogeneous) system. In practice, however, it is far from obvious how to derive analytical results that simultaneously encode scaling and interfacial properties, which are related to short and large distance effects, respectively. 

It has been recently shown \cite{DSS} how the problem can be dealt with within the particle description of field theory. Indeed, the bulk field theory possesses a complete basis of particle states that allow to write the configurational sums in momentum space, and this also in the case of boundary conditions that induce the presence of an interface. The required condition $R\gg\xi$ then projects the calculation to a low energy limit in which the geometry of the system plays a main role. The interface and its fluctuations emerge as due to the propagation of particle modes distributed along a string with a density related to the interfacial tension. At the same time, the dependence on critical exponents is automatically encoded. In particular, the mass of the particle modes coincides with the inverse correlation length, and scales for $T\to T_c$ with the exponent $\nu$. 

In this paper we show how the formalism extends to the case of an interface whose fluctuations are constrained by the presence of a wall. We show how the presence of the wall is implemented in momentum space and how it affects the dependence on the distance from the wall of the order parameter profile resulting from the fluctuations of the interface. We then consider the case in which the tuning of a boundary field can induce the binding of the interface to the wall, and show that the particle formalism naturally accounts for the binding transition and its key parameter.

The paper is organized as follows. In the next section we show how the problem of the interface in presence of the wall is implemented starting from the particle modes of the bulk field theory. In section~3 we use the formalism to determine the order parameter profile and the passage probability of the interface. Section~4 is then devoted to the binding transition induced by a sufficiently attractive wall-interface interaction, while section~5 contains some concluding remarks.

\section{Interface in presence of a wall}
The universal properties of an interface in presence of a wall that we consider in this paper find their simplest implementation within the three-dimensional Ising model defined by the reduced Hamiltonian
\be
\mathcal{H}=-\frac{1}{T} \sum_{\langle i,j\rangle} s_i s_j\,,
\label{lattice}
\ee
where the spin variable located at site $i$ of a cubic lattice takes the values $s_i=\pm 1$, and $\langle i,j\rangle$ means that the sum is performed over all pairs of nearest-neighbor sites. We consider values of the temperature $T$ below the critical value $T_c$, namely in the regime in which the spin reversal symmetry of the Hamiltonian (\ref{lattice}) is spontaneously broken and the absolute value of the magnetization is $|\langle s_i\rangle|=M>0$, where $\langle\cdots\rangle$ denotes the average over all spin configurations weighted by $e^{-{\cal H}}$. More precisely, we consider temperatures only slightly below $T_c$, in such a way that the large correlation length (it diverges as $\xi\simeq|T-T_c|^{-\nu}$ as $T\to T_c$) allows to take the continuum limit. The latter defines an Euclidean (translationally and rotationally invariant in the three dimensions) field theory that we call the bulk field theory \cite{ID,Zinn}. This Euclidean field theory can also be seen as the analytic
  continuation to imaginary time of a quantum field theory defined in two space and one time dimensions. Denoting by $r=(x,y,z)$ a point in Euclidean space, we will identify $z$ as the imaginary time direction. In the continuum the discrete spin variables $s_i$ are replaced by the spin field $s(r)$. 

\begin{figure}[t]
\begin{center}
\includegraphics[width=8cm]{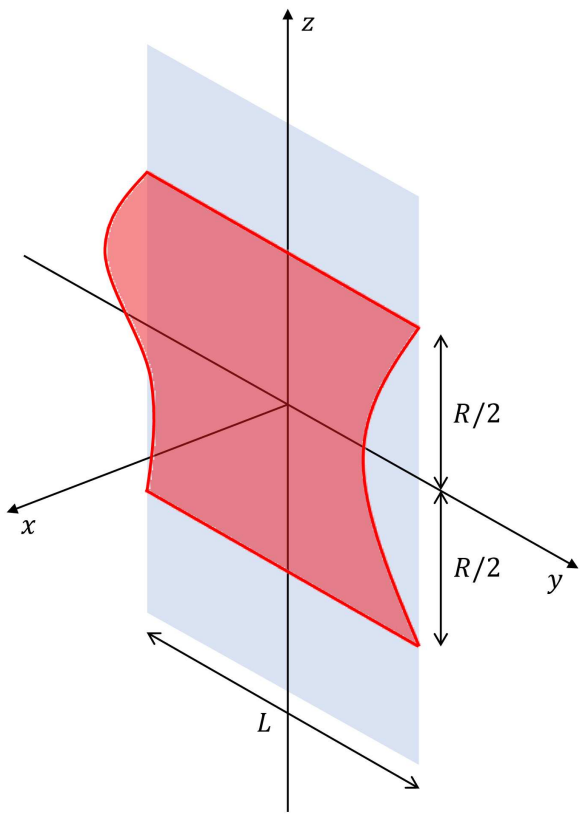}
\caption{Geometry considered in the theoretical derivation, with $L\to\infty$ and $R$ much larger than the bulk correlation length $\xi$. One configuration of the interface is shown.}
\label{geometry}
\end{center}
\end{figure}

In order to study the interfacial problem of our interest, we consider the system in the half-volume $x\geq 0$, with the spins on the wall $x=0$ fixed to the values $s_i=1$ for $|z|< R/2$ and $s_i=-1$ for $|z|>R/2$, where $R$ is much larger than the bulk correlation length $\xi$. Denoting by $\langle\cdots\rangle_{+-}$ configurational averages with these boundary conditions, it is clear that $\lim\limits_{x\to +\infty}\langle\sigma(x,y,0)\rangle_{+-}$ is $-M$ for $R$ finite and $M$ for $R$ infinite. Hence, one expects the presence of an interface pinned along the boundary-condition-changing lines $z=\pm R/2$ on the wall (figure~\ref{geometry}), separating an inner phase with positive magnetization from an outer phase with negative magnetization, and whose average distance from the wall at $z=0$ diverges with $R$. In the following we will show how this result indeed emerges within the field theoretical description of the problem. 

For this purpose, we recall that the bulk field theory admits a particle description. The particles correspond to the excitations modes with respect to the ground state (vacuum) of the quantum field theory, and should not be confused with the molecules of a fluid. Since the rotational invariance of the bulk Euclidean theory is mapped into relativistic invariance of the quantum theory in (2+1) dimensions, the energy $E_{\bf p}$ of a particle mode with mass $m$ and momentum ${\bf p}=(p_x,p_y)$ obeys the relativistic dispersion relation $E_{\bf p}=\sqrt{{\bf p}^2+m^2}$. A complete basis onto which generic excitations can be expanded is provided by the asymptotic $n$-particle states $|{\bf p}_1,{\bf p}_2,\ldots,{\bf p}_n\rangle$ of the bulk theory (see e.g. \cite{Ryder} for an introduction). These states are eigenstates of the energy and momentum operators $H$ and ${\bf P}$ of the quantum theory, with eigenvalues $\sum_{i=1}^nE_{{\bf p}_i}$ and $\sum_{i=1}^n{{\bf p}_i}$, respectively. Th
 e operators $H$ and ${\bf P}$ also act as generators of space-time translations, and for a generic field $\Phi(r)$ we have
\EQ
\Phi(r)=e^{ixP_x+iyP_y+zH}\,\Phi(0)\,e^{-ixP_x-iyP_y-zH}\,.
\label{translations}
\EN

In field theory, interfaces\footnote{As well as other inhomogeneities, see \cite{DSS2}.} are produced by the propagation of particles between the pinning points \cite{DV,DS_bubbles,wedge,DSS}, in the present case the lines $z=\pm R/2$ at $x=0$. Translation invariance in the $y$ direction implies that the number $N$ of propagating particles is extensive in that direction, and is therefore infinite. In order to regulate our expressions, we will denote by $L$ the size of the system in the $y$ direction, always understanding that $N\propto L\to\infty$. The interface is then spanned by the propagation in the imaginary time direction $z$ of an excitation (a string) containing $N/L$ particles per unit length. The propagation occurs between states $|B(\pm R/2)\rangle=e^{\pm\frac{R}{2}H}|B(0)\rangle$ that we can expand over the basis of particle states of the bulk theory in the form
\begin{equation}
|B(\pm R/2)\rangle = \frac{1}{\sqrt{N!}}\, \int \prod\limits^{N}_{i=1}\frac{d\textbf{p}_i}{(2\pi)^2\, E_{\textbf{p}_i}}\, f(\textbf{p}_1,..., \textbf{p}_N)\, e^{\pm R/2 \sum\limits^N_{i=1} E_{\textbf{p}_i}}\, \delta\left(\sum\limits^N_{i=1}p_{y,i}\right)\, |\textbf{p}_1,...,\textbf{p}_N\rangle + ...\,,
\label{boundary_states}
\end{equation}
where $f(\textbf{p}_1,...,\textbf{p}_N)$ is an amplitude, the delta function enforces translation invariance in the $y$ direction, and the state normalization $\langle \textbf{p}|\textbf{q}\rangle=(2\pi)^2\, E_{\textbf{p}}\, \delta(\textbf{p}-\textbf{q})$ is adopted. For reasons that will become clear in a moment, the contribution that we write explicitly in (\ref{boundary_states}) is that of the particles with the lowest mass. The latter is denoted by $m$ and determines the large distance decay of the bulk spin-spin correlation function as $\langle s(r)s(0)\rangle\sim e^{-m|r|}$, a relation that implies
\EQ
\xi=1/m\,.
\label{mass}
\EN

Correlation functions of fields located in the region $|z|<R/2$ of the system with the interface will be computed between the states $|B(-R/2)\rangle$ and $|B(R/2)\rangle$. It follows that the partition function is given by
\begin{align}
Z_{+-}&=\langle B(R/2)|B(-R/2)\rangle= \langle B(0)| e^{-R\, H}|B(0)\rangle\nonumber\\
&\sim \frac{L}{2\pi}\,  \int\prod^N_{i=1}\frac{d\textbf{p}_i}{(2\pi)^2\, m}\, |f(\textbf{p}_1,...,\textbf{p}_N)|^2\, \delta\left(\sum\limits^N_{i=1}p_{y,i}\right)\,e^{-R\left(Nm+\sum\limits_{i=1}^N\frac{{\bf p}_i^2}{2m}\right)}\,,
\label{partition_function}
\end{align}
where in the last line we exploited the fact that the limit of large $R$ forces all momenta to be small, and used the regularization $\delta(0)=L/2\pi$ following from $2\pi\delta(p)=\int e^{ipy}dy$. Here and in the following the symbol $\sim $ indicates omission of terms subleading for large $R$. 

So far we took into account that the interface runs between the pinning axes, but not the presence at $x=0$ of a wall that the interface cannot cross. This information has to be carried by the function $f(\textbf{p}_1,...,\textbf{p}_N)$, which plays the role of emission/absorption amplitude of the particles at the pinning axes. We then impose that none of the particles stays at $x=0$, namely that $f(\textbf{p}_1,...,\textbf{p}_N)$ vanishes when at least one of the momentum components $p_{x,i}$ vanishes. Taking into account that the particles in (\ref{boundary_states}) play a symmetric role, and that $f(\textbf{p}_1,...,\textbf{p}_N)$ should be analytic in the limit of small momenta required for the calculations at large $R$, we write
\begin{equation}
f(\textbf{p}_1,...,\textbf{p}_N)\simeq f_0\, \prod^N_{i=1} p_{x,i}\,, \qquad\textbf{p}_i,\ldots,{\bf p}_N\to 0\,,
\label{condition}
\end{equation}
where $f_0$ is a constant. Plugging this expression in (\ref{partition_function}) we obtain
\begin{equation}
Z_{+-}\sim \frac{L\, |f_0|^2\, e^{-RNm}}{(2\pi)^{2(N+1)}}\, \left(\frac{2\pi m}{R^2}\right)^N\, \sqrt{\frac{2\pi R}{N m}}\,.
\label{result_partition_function}
\end{equation}
This result shows, in particular, how a state with a particle of mass $m$ replaced by one of mass $m'>m$ contributes to the large $R$ expansion a term further suppressed by a factor $e^{-(m'-m)R}$.

The interfacial tension $\sigma$ is the free energy per unit area contributed by an interface whose size is infinite in both the $y$ and the $z$ directions. Since the limit $L\to\infty$ of the size in the $y$ direction is understood, we have
\EQ
\sigma= -\lim_{R\to\infty}\frac{1}{LR}\ln Z_{+-}= \kappa\,m^2=\frac{\kappa}{\xi^2}\,,
\label{tension}
\EN
where
\EQ
\kappa= \frac{N\xi}{L}\,
\label{kappa}
\EN
is dimensionless, and then universal. Since we expect and will show in the next section that the average distance of the interface from the wall increases with $R$, the presence of the wall does not affect the interfacial tension. Hence, the Monte Carlo determination $\sigma\xi^2=\kappa=0.1084(11)$ obtained for the three-dimensional Ising model in absence of the wall \cite{CHP} continues to hold. It follows that the average interparticle distance $L/N=\xi/\kappa$ in the $y$ direction is approximately ten correlation lengths, meaning that the interparticle interaction is negligible. This is nicely consistent with our finding that in the large $R$ limit the particle propagation between the pinning axes is only subject to translation invariance in the $y$ direction (delta function in (\ref{boundary_states})) and to the presence of the wall (expression (\ref{condition})).

\section{Order parameter profile}
The expectation value of a field $\Phi(r)$ at $z=0$ is given by
\begin{align}
G_{\Phi}(x)&\equiv\langle \Phi(x,y,0)\rangle_{+-}= \frac{1}{Z_{+-}}\, \langle B(R/2)|\Phi(x,y,0)|B(-R/2)\rangle \nonumber\\
&\sim \frac{|f_0|^2\, e^{-RNm}}{Z_{+-} N!}\, \int \prod\limits^N_{i=1}\left(\frac{d \textbf{p}_i}{(2\pi)^2\, m}\,  \frac{d  \textbf{q}_i}{(2\pi)^2\, m}\,p_{x,i}\,q_{x,i}\right)\, \delta\left(\sum\limits^N_{i=1} p_{y,i}\right)\, \delta\left(\sum\limits^N_{i=1}q_{y,i}\right)\nonumber\\
&\times e^{-\frac{R}{4m}\, \sum\limits^N_{i=1} (\textbf{p}_i^2+\textbf{q}_i^2)+ix\, \sum\limits^N_{i=1} (p_{x,i}-q_{x,i})}\, F_{\Phi}(\textbf{p}_1,...,\textbf{p}_N|\textbf{q}_1,...,\textbf{q}_N)\,,
\label{vPhi0}
\end{align}
where we used (\ref{translations}) to extract the coordinate dependence, the large $R$ limit has again been taken, and the matrix element
\bea
&& \hspace{-.5cm}F_\Phi({\bf p}_1,\ldots,{\bf p}_N|{\bf q}_1,\ldots,{\bf q}_N)= \langle{\bf p}_1,\ldots,{\bf p}_N|\Phi(0)|{\bf q}_1,\ldots,{\bf q}_N\rangle\label{ff}\\
&& \hspace{-.5cm}= \langle{\bf p}_1,\ldots,{\bf p}_N|\Phi(0)|{\bf q}_1,\ldots,{\bf q}_N\rangle_c+(2\pi)^2m\,\delta({\bf p}_1-{\bf q}_1)\langle{\bf p}_2,\ldots,{\bf p}_N|\Phi(0)|{\bf q}_2,\ldots,{\bf q}_N\rangle_c+\ldots\nonumber
\eea
is evaluated for small momenta. The second equality expresses the decomposition of the matrix elements into connected (subscript $c$) and disconnected parts produced by annihilations of particles on the left with particles on the right \cite{Ryder}; the dots indicate that one has to take into account all possible annihilations. Since the form of (\ref{vPhi0}) implies that each power of momentum contributes a factor $R^{-1/2}$, and each annihilation in (\ref{ff}) yields a delta function $\delta({\bf p}_i-{\bf q}_j)$, and then a factor $R$, the leading contribution to (\ref{vPhi0}) for large $R$ is produced by the maximal number of annihilations. On the other hand, $N$ annihilations just leave a constant $C_\Phi$, so that the leading $x$-dependence is obtained from $N-1$ annihilations, which can be performed in $N!N$ ways. Taking all this into account, we arrive at the expression
\begin{equation}
G_{\Phi}(x)\sim C_{\Phi}+\frac{\kappa\, R^2}{(2\pi)^2\, m^2}\, \int d\textbf{p}\, d\textbf{q}\, p_x\, q_x\,\delta(p_y-q_y)\, F^c_{\Phi}(\textbf{p}|\textbf{q})\, e^{-\frac{R}{4m}\, (\textbf{p}^2+\textbf{q}^2)+i\, x\, (p_x-q_x)}\,,
\label{Gs}
\end{equation}
where $F^c_{\Phi}(\textbf{p}|\textbf{q})\equiv\langle \textbf{p}|\Phi(0)|\textbf{q}\rangle_c$. In particular, we see that, if $F^c_\Phi(\textbf{p}|\textbf{q})$ behaves for small momenta as momentum to the power $\alpha_\Phi$, $G_\Phi(x)-C_\Phi$ will behave as $R^{-(1+\alpha_\Phi)/2}$.

The matrix elements (\ref{ff}) refer to the bulk theory and do not depend on the geometry considered for the interfacial problem. For the spin field $s(r)$ the functional form
\begin{equation}
F^c_s(\textbf{p}|\textbf{q})|_{p_y=q_y}=\frac{c_s}{p_x-q_x}\,, \qquad p_x,q_x\to 0\,
\label{matrix_element_Fs}
\end{equation}
was deduced in \cite{DSS}. When inserting this expression in (\ref{Gs}) it is convenient to get rid of the pole by differentiating with respect to $x$. Performing the momentum integrations and integrating back in $x$ with the boundary conditions $\lim\limits_{x\to+\infty}G_s(x)=-M$ and $G_s(0)=M$ then gives the order parameter (or magnetization) profile
\begin{equation}
G_s(x)\sim M+2M\left[\frac{2}{\sqrt{\pi}}\, \eta\, e^{-\eta^2}-\text{erf}(\eta) \right]\,,
\label{onepoint_magnetization}
\end{equation}
with
\begin{equation}
\eta=\sqrt{\frac{2}{R\xi}}\,x\,
\label{eta}
\end{equation}
and $c_s=4iM/\kappa$. Using (\ref{translations}) the calculation can be straightforwardly extended to a generic $z\in(-R/2,R/2)$. The effect is that in (\ref{onepoint_magnetization}) $\eta$ is replaced by $\chi=\eta/\sqrt{1-(2z/R)^2}$.

\begin{figure}[t]
\begin{center}
\includegraphics[width=12cm]{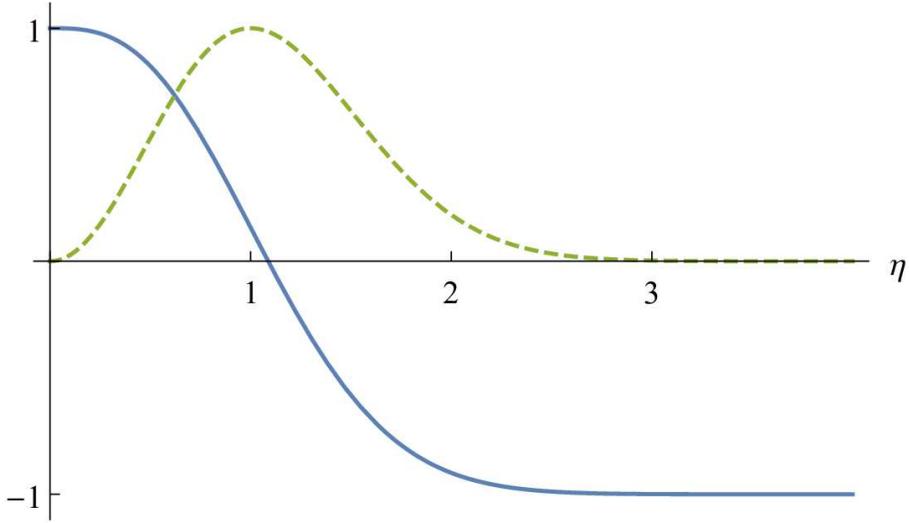}
\caption{Order parameter profile $G_s(x)/M$ (Eq.~(\ref{onepoint_magnetization}), continuous curve) and passage probability $p(x)/p(\sqrt{R\xi/2})$ (Eq.~(\ref{passage_pdf}), dashed curve).}
\label{profiles}
\end{center}
\end{figure}

The result (\ref{onepoint_magnetization}) admits a simple probabilistic interpretation once we look at this leading contribution in the large $R$ expansion as due to an interface that sharply separates two pure phases. Then the magnetization at a point $r=(x,0,0)$ within a configuration in which the interface intersects the $x$-axis at a point $u$ can be written as
\begin{equation}
s(x|u)=M\, \theta(u-x)-M\, \theta(x-u)\,,
\label{spin_x|u}
\end{equation}
where $\theta(x)$ is the step function that vanishes for $x<0$ and equals 1 for $x>0$. If $p(u)\, du$ is the probability that the interface intersects the $x$-axis in the interval $(u,u+du)$, then the average magnetization can be written as
\begin{equation}
\overline{s(x)}=\int_0^{+\infty}du\, p(u)\, s(x|u)=M\int_x^{+\infty}du\, p(u)-M\int_0^x du\, p(u)\,.
\end{equation}
This expression coincides with (\ref{onepoint_magnetization}) for a passage probability density
\begin{equation}
p(x)=4 \sqrt{\frac{2}{\pi R\xi}}\, \eta^2\, e^{-\eta^2}\,,
\label{passage_pdf}
\end{equation}
which correctly satisfies $p(x)\geq 0$ and $\int^{+\infty}_0 dx\, p(x)=1$. $p(x)$ is maximal at $\eta=1$ (figure~\ref{profiles}), showing that the average distance of the interface from the wall increases as $\sqrt{R}$. In addition, $p(0)=0$ verifies in real space the impenetrability of the wall that we imposed in momentum space through the condition (\ref{condition}).

The probabilistic interpretation also illustrates that the fluctuations of the interface in the $y$ direction do not affect the leading term of the local magnetization in the large $R$ expansion. Then it is not surprising that the profile (\ref{onepoint_magnetization}) is analogous to that known in two dimensions \cite{AI,Abraham_review,DS1}, i.e. in absence of the $y$ direction\footnote{The fluctuations in the $y$ direction should show up at leading order in the large $R$ expansion of the spin-spin correlation function, which in two dimensions was obtained in \cite{long_range}.}. It must be noted, however, that the factor $\sqrt{2}$ in (\ref{eta}) is absent in two dimensions. This is due to the fact that in two dimensions the particle modes of the Ising model below $T_c$ are topological excitations (they are kinks, see e.g. \cite{Ryder}). Since the spin field is topologically neutral, the lightest state to which it couples is a two-particle (kink-antikink) one (see \cite{McW} and, f
 or a review, \cite{immf}). It follows that in two dimensions the relation (\ref{mass}) is replaced by $\xi=1/2m$, and this difference propagates in the results expressed in terms of the correlation length.

It must also be observed that the impenetrability of the wall is the only boundary effect that we took into account in our theoretical derivation. In actual measurements (in particular in simulations on the lattice) the value of the order parameter close enough to the wall will be affected by the specific nature of the interaction between the wall and the bulk degrees of freedom. Hence, the results (\ref{onepoint_magnetization}) and (\ref{passage_pdf}) hold for $x$ larger than few correlation lengths. Since the main interfacial effects occur around $x\propto\sqrt{R}$, and $R\gg\xi$, they are not affected by boundary details, unless we consider the generalization of the next section.

\section{Binding transition}
The system settings considered so far lead to an interface whose average distance from the wall diverges as $\sqrt{R}$. On the other hand, the introduction of a tunable boundary field can lead to a wall-interface interaction sufficiently attractive to determine a binding of the interface to the wall. Conversely, the passage from the binding to the fluctuating regime corresponds to a transition that is most often referred to as "wetting" transition (see \cite{deGennes,Dietrich,BEIMR} for reviews). This terminology refers to a liquid-vapor interface, the liquid phase being that in contact with the wall. 

As we now explain, the particle formalism naturally accounts also for the binding transition. We saw that in the limit relevant for phase separation (linear size of the interface much larger than the bulk correlation length $\xi$) the interfacial properties are determined by low energy particle modes whose mutual interaction is negligible due to a large average separation. The interaction of a particle with the wall can be characterized within the scattering framework, in which an incoming particle has momentum ${\bf p}=(p_x<0,p_y)$. At low energy the interaction with the wall is elastic and the particle bounces back with momentum ${\bf p}=(-p_x,p_y)$, the component $p_y$ being conserved due to translation invariance in the $y$ direction. The relation $E^2={\bf p}^2+m^2$ defines the parameter $\beta$ such that
\bea
&& E=m\cosh\beta\,,\\
&& |{\bf p}|=m\sinh\beta\,.
\label{pbeta}
\eea 
If the particle-wall interaction is sufficiently attractive, the particle will bind to the wall and, as usual in scattering theory \cite{Landau,ELOP}, the bound state corresponds to a value $E_0<m$ of the energy, namely to ${\bf p}^2<0$, or $\beta=i\theta_0$ with $\theta_0\in(0,\pi)$. It follows that in the bound regime the contribution of the interface to the energy per unit length is $\frac{N}{L}m\cos\theta_0=\sigma\cos\theta_0$, where we used (\ref{tension}) and (\ref{kappa}). Hence, if $e$ is the energy per unit length associated to the wall, the energy per unit length of the wall-interface bound state is
\EQ
\tilde{e}=e+\sigma\cos\theta_0\,.
\label{young}
\EN
The value of the binding angle $\theta_0$ depends on the strength of the particle-wall interaction, and the unbinding transition occurs at $\theta_0=0$, when the binding energy per unit length $\sigma(1-\cos\theta_0)$ vanishes. 

\begin{figure}[t]
\begin{center}
\includegraphics[width=4cm]{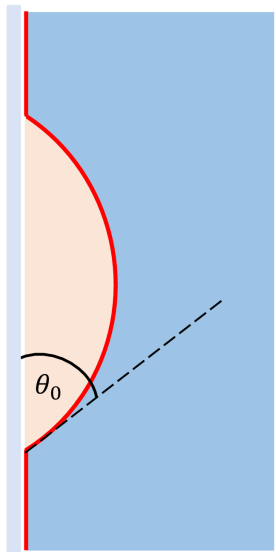}
\caption{In field theory a drop on the wall corresponds to the unbinding and recombination of a wall-interface bound state. The contact angle $\theta_0$ vanishes at the unbinding transition.}
\label{contact_angle}
\end{center}
\end{figure}

Remarkably, (\ref{young}) accounts for the basic relation of the phenomenological wetting theory \cite{deGennes}, namely the equilibrium condition for a liquid drop on the wall, in which $\tilde{e}$ and $e$ are the wall-vapor and wall-liquid surface tensions, respectively, and $\theta_0$ is the angle that the drop forms with the wall (figure~\ref{contact_angle}). The wetting transition occurs at $\theta_0=0$, when the drop spreads on the wall. 

Consider now the dependence of $\theta_0$ on the parameters of the system. The wall contributes to the Euclidean action of the theory a term $h\int dydz\,\Phi_B(0,y,z)$. Since the action is dimensionless, if $X_B$ is the scaling dimension of the boundary field $\Phi_B$, the coupling $h$ has the dimension of a mass (or inverse length) to the power $2-X_B$. Hence, $\theta_0$ is a function of the dimensionless combination $h/m^{2-X_B}$, where $m=1/\xi\sim (T_c-T)^\nu$. For $h$ fixed, the condition $\theta_0=0$ determines the unbinding (or wetting) transition temperature $T_w(h)<T_c$. It is clear that for $T$ sufficiently close to $T_c$, namely for a mass $m$ sufficiently small, the near-critical fluctuations become too strong and the particles have to be unbounded, so that the bound regime corresponds to $T<T_w$. We also see that the interface is unbound for $h=\infty$, which corresponds to the boundary field considered in the previous sections. 

It is customary (see \cite{Dietrich,BEIMR}) to characterize the transition through the exponent $\alpha_S$ defined for $T\to T_w^-$ by
\EQ
(1-\cos\theta_0)\propto(T_w-T)^{2-\alpha_S}\,,
\label{alpha}
\EN
and the transition is said to be continuous if $\alpha_S<1$. The terminology refers to the continuity of the first derivative of (\ref{alpha}) at $T_w$, taking into account that the contact angle $\theta_0$ is phenomenologically set to zero in the unbound regime $T_w<T<T_c$. We can get insight on the exponent $\alpha_S$ recalling that, as usual in scattering theory, the bound state corresponds to a pole at $E=E_0$ in the scattering amplitude of the particle on the wall. Then general analytical properties of the amplitude \cite{Landau,ELOP} tell us that when we move from the bound to the unboud regime, namely when $T$ increases through $T_w$, the pole does not disappear, but slides through a square root branch point at $E=m$ into a second sheet of the complex energy plane. Within the parametrization $E=m\cos\theta_0$ this corresponds to a continuation from positive to negative values\footnote{Such a continuation is regularly exploited in the context of exact scattering solutions, see 
 \cite{GZ}.} of $\theta_0$, namely to
\EQ
\theta_0\propto(T_w-T)^{2n+1}\,,\hspace{.7cm}n=0,1,2,\ldots\,
\label{wetting}
\EN
in the vicinity of $T_w$. Comparison with (\ref{alpha}) then yields
\EQ
\alpha_S=-4n\,.
\EN
Clearly, the generic case is expected to correspond to $n=0$, and then to $\alpha_S=0$. As reviewed in \cite{BEIMR}, this value agrees with numerical simulations within the Ising model\footnote{We also notice that the value $\alpha_S\approx -5$ deduced from a phenomenological renormalization group approach (see \cite{BEIMR} and references therein) is reminiscent of the case $n=1$, i.e. $\alpha_S=-4$.} \cite{Binder}.  

A second exponent $\beta_S<0$ describes the divergence of the distance of the interface from the boundary,
\EQ
l\propto(T_w-T)^{\beta_S}\,,
\label{beta}
\EN 
as $T\to T_w^-$. In the scattering framework $l$ is related to the decay $e^{-x/l}$ of the wave function for a distance $x\to\infty$ from the wall in the bound regime. Such a behavior can be seen as originating from the plane wave $e^{ip_x x}$ and the imaginary value of the momentum in the bound regime: $|{\bf p}|=im\sin\theta_0$ from (\ref{pbeta}). Close to $T_w$, where $\theta_0$ is small, one could naively argue $l\propto 1/m\theta_0$, and infer $\beta_S=\alpha_S/2-1$ from  comparison with (\ref{alpha}) and (\ref{beta}). $\alpha_S=0$ then leads to $\beta_S=-1$, a value that has been observed experimentally \cite{Ragil}. However, experimental systems include long range interactions that are not present in our framework. The safest comparison is that with simulations within the nearest-neighbor Ising model\footnote{It can be noted that the value $\alpha_S=0$, which agrees with simulations in the Ising model, is also consistent with the experiment of \cite{RBM}.}, which are consisten
 t with $l\propto|\ln(T_w-T)|$ (see \cite{Binder} and the discussion in \cite{PE}). This indicates that the fact that $p_x\to 0$ does not imply $|{\bf p}|\to 0$ cannot be forgotten. The implication holds instead in two dimensions \cite{localization} (i.e. in absence of the $y$ direction), where the values $\alpha_S=0$ and $\beta_S=-1$ indeed correspond to the exact Ising lattice solution of \cite{Abraham_review,Abraham_wetting}.

\section{Conclusion}
In this paper we have shown how the universal properties of phase separation in presence of a wall in three dimensions can be derived in terms of the particle modes that are the elementary degrees of freedom of the bulk field theory. The interface emerges as due to the propagation in imaginary time of particles distributed along a string. We implemented the presence of the wall within the configurational sum in momentum space and showed how this leads to the expected properties of the passage probability of the interface in coordinate space. The theory relates the interfacial tension to the particle density along the string, and shows how the propagation of the particles between the pinning axes is affected by the presence of the wall, while the interaction among the particles is negligible due to a large average interparticle distance. 

We also showed that the particle formalism naturally describes, via scattering theory, the transition to the regime in which the interface is bound to the wall by a sufficiently strong attractive interaction. The temperature dependence of the wall-interface binding energy is carried by a parameter $\theta_0$ that originates from the relativistic dispersion relation of the particle modes and finally provides the contact angle of phenomenological wetting theory.

%\end{document}
%\newpage

%\vspace{1cm}
%\noindent \textbf{Acknowledgments.} 

\end{document}